\documentclass[useAMS,usenatbib]{mn2e}
\usepackage{graphicx}

\newcommand{\revised}[1]{{#1}}

\title[A 34 Hz QPO in GRS 1915+105]{Discovery of a 34 Hz Quasi-Periodic Oscillation in the X-ray emission of GRS 1915+105}
\author[T. M. Belloni, D. Altamirano]{T.M. Belloni$^{1}$\thanks{E-mail:
tomaso.belloni@brera.inaf.it}, D. Altamirano$^{2}$\\
$^{1}$INAF - Osservatorio Astronomico di Brera, Via E. Bianchi 46, I-23807, Merate, Italy\\
$^{2}$Astronomical Institute ``A. Pannekoek", University of Amsterdam, Science Park 904, 1098 XH Amsterdam, The Netherlands}
\begin{document}

\date{Accepted 2013 February 12. Received 2013 February 12; in original form 2012 December 13}

\pagerange{\pageref{firstpage}--\pageref{lastpage}} \pubyear{2013}

\maketitle

\label{firstpage}

\begin{abstract}
We report the discovery in the Rossi X-Ray Timing Explorer data of GRS
1915+105 of a second quasi-periodic oscillation at 34 Hz, simultaneous
with that observed at 68 Hz in the same observation. The data
corresponded to those observations from 2003 where the 68-Hz
oscillation was very strong. The significance of the detection is
4.2$\sigma$.  These observations correspond to a very specific
position in the colour-colour diagram for GRS 1915+105, corresponding
to a harder spectrum compared to those where a 41 Hz oscillation was
discovered. We discuss the possible implications of the new pair of frequencies
comparing them with the existing theoretical models.

\end{abstract}


\section{Introduction}

High-Frequency Quasi-Periodic Oscillations (HFQPO) are signals at
frequencies 30-450 Hz observed in the X-ray light curves of black-hole
binaries (BHBs). These correspond to the time scales expected from
Keplerian motion in the innermost regions of the gravitational well of
the black hole. They constitute the most promising way to use timing
signals to measure fundamental parameters of the black hole such as
its spin and the existence of a innermost stable orbit Unfortunately,
their investigation has been hindered by the scarcity of detections
available from 16 years of operations of the Rossi X-Ray Timing
Explorer (RXTE), whose Proportional Counter Array (PCA) was the only
instrument able to provide the necessary signal to noise for the
detections. Very few detections were reported from thousands of
high-quality observations of BHBs (see Belloni, Sanna \& M\'endez 2012
and references therein; see also Altamirano \& Belloni 2012 on the
discovery of 67 Hz HFQPOs in the black hole system IGR J17091--3624;
Altamirano et al. 2011). There is evidence that these oscillations
appear at specific frequencies for each source, although the small
number of cases does not allow to say something more firm. In a few
cases, pairs of peaks have been observed simultaneously. In these
cases, the peaks appear to be close to small integer ratios (2:3,
3:5). This has been interpreted associating the QPO frequencies to
relativistic time scales at a specific radius, where these frequencies
are in resonance, resulting therefore in special frequency ratios
(Klu\'zniak \& Abramowicz 2001).  Another model, the relativistic
precession model (see Stella et al. 1999 and references therein),
makes predictions about the frequencies of both high-frequency peaks,
but does not lead to specific ratios. Other models exist, which
predict different spectra of signals potentially observable in the
data.

One system stands out in this and many other respects: the very bright
BHB GRS 1915+105 (see Fender \& Belloni 2004 for a review). This very
peculiar BHB has been observed thousands of times by RXTE. A few cases
of HFQPOs
from GRS 1915+105 were reported in the literature (Morgan et al. 1997;
Strohmayer 2001; Belloni et al. 2001; Remillard et al. 2003; Belloni
et al. 2006). Strohmayer (2001) reported the discovery of two
simultaneous oscillations at 41 and 69 Hz (which give the 2:3 ratio
mentioned above).  Recently, Belloni \& Altamirano (2012, hereafter
BA12) have carried out a systematic study of all RXTE observations of
GRS 1915+105 and have detected HFQPOs in \revised{51} observations. All but
three of these were restricted to a small range of frequencies between
63 and 71 Hz. Their procedure did not lead to the detection of
pairs. However, Strohmayer (2001) averaged several observations to
reach his detections.  Here we report the detection of an additional
HFQPO at 34 Hz, observed when the higher-frequency one was around 68
Hz\revised{. We obtained it by} averaging the observations corresponding to 3 of the 5
strongest detections in BA12.

\section[]{Data analysis}

\begin{figure}
\begin{center}
\includegraphics[width=8.5cm]{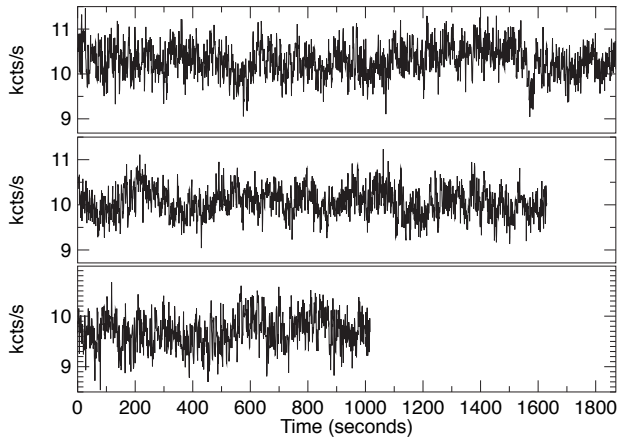}
\end{center}
\caption{Light curves for the three observations (resolution 1
  second, 2-15 keV). From top to bottom observations 80701-01-28-00,
  80701-01-28-01, 80701-01-28-02.}
\label{fig:licu}
\end{figure}

In the sample of HFQPOs of BA12, only five cases featured a detection
with a single-trial significance larger than 10$\sigma$. Two of them
correspond to early 1996 observations (ObsIDs 10408-01-04-00/06-00,
corresponding to Obs. \#3 and \#5 in BA12, respectively) and are the
same as those reported by Morgan et al. (1997). The remaining three
(ObsIDs 80701-01-28-00/01/02, corresponding to Obs. \#38, \#39 and
\#40 in BA12, respectively) are observations made on Oct 21, 2003 (MJD
52933), when the HFQPO was at a frequency of 68.3, 67.8 and 67.8 Hz
for the three observations, respectively. The light curves of the
three observations at 1-s time resolution can be seen in
Fig. \ref{fig:licu}. In the classification of Belloni et al. (2000)
they correspond to class $\delta$.
\revised{In comparison, the strong detection obtained in 1996 (Morgan et al. 1997)
corresponded to class $\gamma$. Regardless of the class, the position in the 
colour-colour diagram (and hence spectral shape) is what appears to be associated
to the presence of the QPO (see below).}

For the timing analysis we followed the same procedure as in BA12.  We
produced a combined power density spectra (PDS) of the three
observations of 2003 Oct 21 by dividing the total dataset (selecting
PCA absolute channels 8--79, 3.28--33.43 keV) in 16-s stretches and
averaging the corresponding PDS. The total exposure was 4576 s,
corresponding to 286 stretches.  The time resolution of the data was
4096 points per second, leading to a Nyquist frequency of $\sim$2
kHz. We normalised the PDS according to Leahy et al. (1983) and
rebinned the powers logarithmically so that each frequency bin was
larger than the previous one by $\sim$2\%. We estimated errors on
power values following van der Klis (1988).

As in BA12, we use the 16-s time-resolution Standard 2 mode data to
calculate X-ray colours. We defined the soft colour (HR1) as the count
rate in the 6.0--16.0 keV band divided by the rate in the 2.0--6.0 keV
band and the hard colour (HR2) as the ratio of the count rates in the
16.0--20.0 keV rate divided by the 2.0--6.0 keV rate. For more
information about how the colours were estimated, see BA12.

\begin{figure}
\begin{center}
\includegraphics[width=8.5cm]{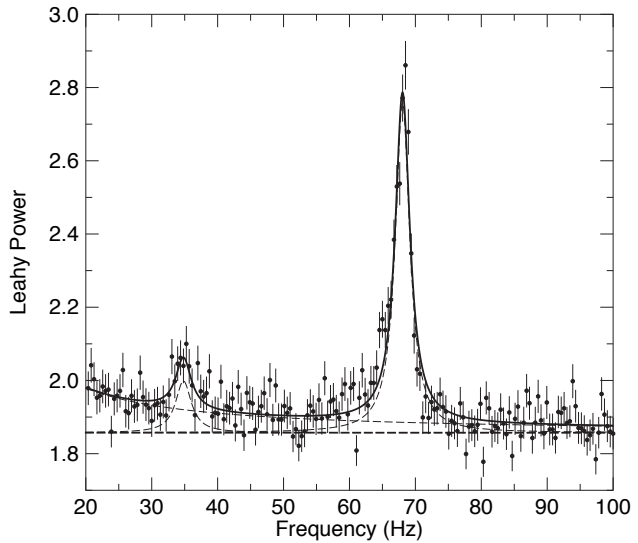}
\end{center}
\caption{PDS of the three combined observations \revised{shown in Fig. 1}.The lines show the
  different components used for the fit and the total model (see text).}
\label{fig:pds}
\end{figure}

\section{Results}

In Fig. \ref{fig:pds} we show the high-frequency part of the PDS
estimated after averaging the three observations performed in 2003 Oct
21. In addition to the strong QPO around 68 Hz, a second peak around
34 Hz is seen. A fit with a model consisting of a flat component to
account for the Poissonian noise, a power law to account for residual
source noise and two Lorentzian components for the two peaks yields a
best fit to the 20-100 Hz region of the PDS with a $\chi^2$ of 190 for
139 degrees of freedom. The relatively high $\chi^2/{\rm dof}$ value
is due to the combination of two facts: the 68 Hz peaks in the three
observations have slightly different centroid frequency (suggesting
that the centroid frequency is drifting moderately with time) and the
peak within one observation is asymmetric. (It is beyond the scope of
this work to understand if the asymmetry we find also within each
observation is intrinsic to the signal or due to frequency drifts
within a $\sim$1000 s).
The best first parameters for the two HFQPOs are shown in
Tab. \ref{tab:parameters}.  The detection significance of the 68 and
34 Hz peaks are 24$\sigma$ and 4.2$\sigma$ respectively (single
trial).  The quality factor (centroid frequency divided by FWHM, i.e.,
a measure of the coherence of a peak in the PDS) is 24.7$\pm$0.3 and
13.1$\pm$1.2, respectively.
\revised{The formal 1$\sigma$ errors on the centroids are very small. However,
additional systematics can affect the accuracy of this measurement. As mentioned above, the peaks are
not symmetric in frequency, due to centroid drifting, although some intrinsic
asymmetry cannot be excluded. Therefore, the statistical accuracy of the centroid
measurements needs to be taken as a lower limit.}
We extracted PDS from energy sub-selections in order to study the
rms spectrum of the 34 Hz QPO, but the signal is too weak to yield
significant results (for the rms spectrum of the 68 Hz QPO, see BA12).

\begin{table}
 \centering
 \caption{Best-fit parameters for the two HFQPOs
 }
  \begin{tabular}{@{}lcc@{}}
  \hline
   Parameter   &
   QPO$_1$    &
   QPO$_2$    \\
 \hline

Centroid (Hz) & 34.96$\pm$0.28& 68.03$\pm$0.06\\
FWHM (Hz)   &   2.67$\pm$0.89&   2.75$\pm$0.16\\
\% rms           &  0.81$\pm$0.09 &   2.04$\pm$0.04 \\
 
\hline
\end{tabular}
\label{tab:parameters}
\end{table}

\begin{figure}
\begin{center}
\includegraphics[width=8.5cm]{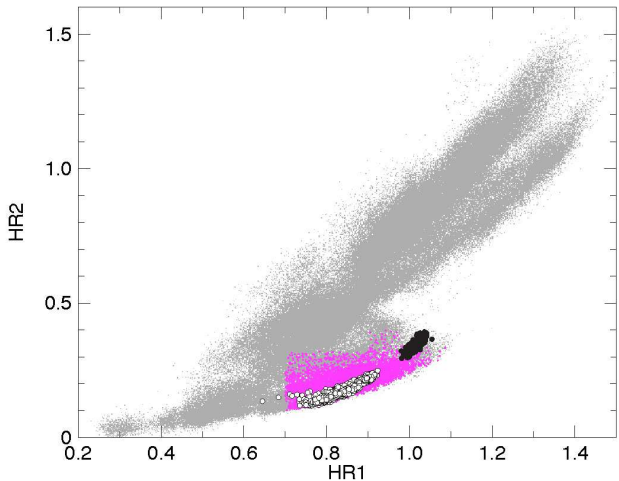}
\end{center}
\caption{Colour-colour diagram for all RXTE/PCA observations of GRS
  1915+105. Each point is accumulated for 16 seconds. All segments
  included in the observations where a HFQPO was detected are marked
  in dark grey (magenta in the colour version, see also BA12). The two
  groups of points marked differently correspond to our selection (black, on the
  right) and to the selection where the 41 Hz HFQPO was found (white, on the
  left, Strohmayer 2001).}
\label{fig:ccd}
\end{figure}

In analogy with what done in BA12, we selected from all 16s intervals
from the full set of archival data those corresponding to similar
X-ray colors and/or count rates and produced average power spectra
from them. 
Figure \ref{fig:ccd} shows the colour-colour diagram including all 16
s data-points of the $\sim$1800 observations available (light grey
points). The dark grey (magenta in the colour version) points are
those with positive 68 Hz QPO detections within each observation from
the full sample (see BA12). The black points are those corresponding
to the three observations analysed here, the white points are those from 
Strohmayer (2001), where the 41 Hz QPO was detected.
It is clear that the points corresponding to the 34 Hz detection are
narrowly distributed in an extreme part of the colour-colour diagram. No
significant detection could be obtained from the relatively low number
of points (a total of no more of $\sim600$ s of data) within the black
area after the three observations reported here have been removed.
We also investigated whether the other two observations (ObsIDs
10408-01-04-00/06-00) with highly significant 68 Hz QPO\revised{, added together,} show any
evidence for a QPO at 34 Hz, but found none\revised{, with a 3$\sigma$ upper limit
of 0.30\% both at the same frequency of 2003 and at half the frequency of 1996.} 
Adding all \revised{other} observations
with detections in single observations (see BA12) did not lead either
to any evidence of the 34 Hz QPO, \revised{with a 3$\sigma$ upper limit of 0.16\% at 34 Hz}.

\section{Discussion}

In this work we report the discovery of a $\sim$34 Hz QPO in
observations of the black hole Low-Mass X-Ray Binary GRS 1915+105. This QPO was
detected during three of the 5 observations (in a total of $\sim$1800
observations) where the 68 Hz QPO was more than $10\sigma$
significant. These three observations occupy a narrow and un-sampled
region of the colour-colour diagram (see Figure~\ref{fig:ccd})
suggesting that the appearance (or the strength) of this QPO is
related to the source state. 
The ratio between the centroid frequency of the high-frequency peak
and that of the low-frequency peak we have discovered is
\revised{
R=1.95$\pm$0.05 (3$\sigma$ error), consistent within 3$\sigma$ with a harmonic nature of the two
signals.} This is the first case of two HFQPOs with such a frequency
ratio in GRS 1915+105.  

Strohmayer (2001) found a similar pair, with frequencies 41.5$\pm$0.4
Hz and 69.2$\pm$0.15 Hz, with a ratio of 1.67$\pm$0.01 averaging five
observations from 1997, all corresponding to variability class
$\gamma$, similar to $\delta$.
In Fig.~\ref{fig:ccd}, we mark the points corresponding to when the
$\sim$41 Hz QPO was found with white circles.
Belloni, M\'endez \& S\'anchez-Fern\'andez 2001 reported the discovery
of a 27 Hz QPO, however, this feature was not detected simultaneously
with other peaks.
Comparing the region marked by black circles (where the 34 Hz QPO was
detected) with that of the observations where a 41 Hz QPO has been
found (white circles), we see that the 34 Hz QPO is found when the
spectrum was harder (see Fig. \ref{fig:ccd}).  Since in this state the
energy spectrum is dominated by the thermal disk component, this
suggests that the 34 Hz feature is associated to a higher disk
temperature than the 41 Hz one.

Given the scarce number of detections of these QPOs, it is not
possible to tell whether the 27 Hz, 34 Hz and the 41 Hz are the same
QPO or not. If the 34 Hz and the 41 Hz are produced by the same
mechanism, we find that its frequency have drifted by 16$\pm$1\%,
while the peak frequency of the 68 Hz QPO only by 1.7$\pm$0.2\%, i.e.,
while the upper QPO remained almost constant (see BA12), the lower one
can drift significantly.
This means that neither the ratio nor the difference between the
centroids of the low- and high-frequency peaks (i.e. 34-41 Hz and 68
Hz, respectively) is constant. None of the existing physical models
(see BA12) to our knowledge is able to reproduce a couple of peaks of
which only one changes frequency.

If the 34 Hz and 41 Hz peaks are two distinct features, together with
the 27 Hz peak detected by Belloni, M\'endez \& S\'anchez-Fern\'andez
2001, we are observing different multiple peaks, which provide an
important observational constraint for theoretical models.  
The relativistic precession model interprets the observed frequencies
in terms of orbital, epicyclic and precession frequencies as predicted
by General Relativity.  As such, in its simplest form it predicts only
two peaks in the high-frequency regime and that they naturally vary
together (see Stella et al. 1999).
The parametric resonance is based on the resonance between relativistic frequencies
at a certain radius. As such, it predicts small integer ratios between
HFQPO frequencies of the form $\omega_1$ = 2$\omega_2$/n, but no major
frequency shift (see Klu\'zniak \& Abramowicz 2001). The sequence
(27)-34-41-68 does not yield simple ratios. 
\revised{For a discussion of sub-harmonics of high-frequency oscillations within 
this framework, see Rebusco et al. (2012).}
The disko-seismic global oscillation model (see Nowak \& Wagoner 1993 and
references therein) interprets the observed signals as trapped global oscillation modes
in the accretion flow. It naturally features harmonic structure in the
oscillations, although a precise identification of the frequencies is
not clear.  
Local acoustic models (Lubow \& Pringle 1993) predict a
spectrum of vertical acoustic oscillation modes. At a given radius,
the frequencies of these modes are $\omega_n$=$\omega_k\sqrt{4/3\, n+1}$ 
where $\omega_k$ is the Keplerian orbital
frequency. Simple combinations of the first frequencies of the series
are already not compatible with the observed values.  
The accretion-ejection instability model is based on the Rossby Wave Instability (RWI),
where a wave is trapped in a magnetised accretion disk and has a direct connection
to jet ejection. It also predicts the excitation of
higher modes, whose relative ratio depending on physical parameters
(Tagger \& Verni\'ere 2006). 
In the inner torus oscillation model, based on acoustic oscillations in the 
innermost regions of the accretion flow
(Rezzolla et al. 2003), a harmonic sequence is natural, but the
sequence observed here is not harmonic.  
Finally, the model proposed
by Psaltis \& Norman (2001) assume a transition radius in the accretion disk and 
study its effect as a filter on a spectrum of density fluctuations. The model 
can have multiple modes excited and yield
multiple peaks. All of these models need to be tested against the new
set of frequencies.

Interestingly, the FWHM of the 34 Hz and 68 Hz peaks in our
observations is consistent with being the same. Unlike the case of
low-frequency QPOs (LFQPOs), where it is the Q factor that remains
constant between different peaks (see Rao et al. 2010, but see Ratti,
Belloni \& Motta 2012), if the mechanism that broadens the peaks is
the same, it is consistent with being an amplitude modulation of the
signal. In other words, the width of the peaks, their quasi-periodic nature, 
could be interpreted as a signal which is modulated in amplitude, where both 
HFQPOs are modulated in the same way.
Assuming that the 34 Hz and the 68 Hz are harmonically related leads
to the possibility that the 34 Hz QPO is the fundamental, or that it
is subharmonic of the 68 Hz QPO.
If the 34 Hz is the subharmonic, then the physical mechanism that
produces the HFQPO must be non-linear.
If the 34 Hz QPO is the fundamental, one would need to understand why
the fundamental can be (much) weaker (in terms of fractional rms
amplitude) than the first harmonic (by a factor larger than 3), and in
almost all cases not even be detected. This would point to a physical
mechanism which can be highly symmetric (such the symmetry inherent in
the geometry of a tilted precessing orbit); however, it could also
point to a scenario where the mechanism that sets the frequency of the
oscillation is independent from that which sets its amplitude (e.g.,
Altamirano et al. 2008 and references therein), the later possibly
favouring in this case the higher frequencies.

Very recently Altamirano \& Belloni 2012 reported on the discovery of
67 Hz QPOs in IGR J17091--3624, the first black-hole system to show
strong phenomenological similarities to those seen in GRS 1915+105
(Altamirano et al. 2011). The fact that the HFQPOs frequency in IGR
J17091-3624 matches surprisingly well with that seen in GRS 1915+105
raises questions on the mass scaling of QPOs frequency in these two
systems, as IGR J17091--3624 might be a factor few less massive than
GRS 1915+105 (Altamirano et al. 2011; although note that these
authors' conclusion strongly depend on the distance and mass of IGR
J17091--3624, both completely unconstrained today).
\revised{Of course, the observed frequencies are likely to depend also on the
(unknown) spin of the source, making the comparison even more complex.}
In any case, if the 34 Hz QPO we detect in GRS 1915+105 represents the
fundamental frequency, mass estimates for GRS 1915+105 will change
depending on the theoretical model assumed and generally leading to a
less massive black-hole (e.g, Nowak et al. 1997, Morgan et
al. 1997). The correct identification of the 34 Hz and 68 Hz QPOs has
therefore important implications on works that use the 68 Hz HFQPO in
GRS 1915+105 as a standard measurement to estimate the mass of
Galactic or supermassive black holes (e.g., Middleton \& Done 2010).

Whatever the answer is, the situation of HFQPOs is clearly complex
once sensitive observations allow us to obtain more detections. This
opens the possibility to make major steps forward with the upcoming
Indian X-ray mission ASTROSAT (Agrawal 2006), whose sensitivity above
10 keV is much improved compared to the RXTE/PCA.  If selected by ESA,
the LOFT mission with its much larger effective area will extend the
current sensitivity down by one order of magnitude and will allow the
study of additional weaker modes, providing a full spectrum of
oscillation modes (see Feroci et al. 2010).

\section*{Acknowledgments}

The research leading to these results has received funding from the
European Community Seventh Framework Programme (FP7/2007-2013) under
grant agreement number ITN 215212 ``Black Hole Universe''.

\end{document}